\documentclass[aps,floats,prd,nofootinbib,twocolumn]{revtex4}

\usepackage{amssymb}
\usepackage{amsmath}

\usepackage{graphicx}
\usepackage{graphics}
\usepackage{dcolumn}
\usepackage{color}
\usepackage{fancyhdr} 
\usepackage{graphicx}

\usepackage{subfig}

\usepackage[utf8]{inputenc}

\usepackage[justification=centerlast, format=plain, labelfont=bf]{caption}

\def\VEV#1{\left\langle #1 \right\rangle}

    \newcommand{\be}{\begin{equation}}
  \newcommand{\ee}{\end{equation}}
    \newcommand{\ba}{\begin{align}}
  \newcommand{\ea}{\end{align}}

\begin{document}

\title{21-cm Fluctuations from Charged Dark Matter}

\author{Julian B.~Mu\~noz\footnote{Electronic address: \tt julianmunoz@fas.harvard.edu}
} 
\affiliation{Department of Physics, Harvard University, 17 Oxford St., Cambridge, MA 02138}
\author{Cora Dvorkin\footnote{Electronic address: \tt cdvorkin@fas.harvard.edu}
} 
\affiliation{Department of Physics, Harvard University, 17 Oxford St., Cambridge, MA 02138}
\author{Abraham Loeb\footnote{Electronic address: \tt aloeb@cfa.harvard.edu }
} 
\affiliation{Astronomy Department, Harvard University, 60 Garden St., Cambridge, MA 02138}

\date{\today}

\begin{abstract}
The epoch of the formation of the first stars, known as the cosmic dawn, has emerged as a new arena in the search for dark matter. 
In particular, the first claimed 21-cm detection exhibits a deeper global absorption feature than expected, which could be caused by a low baryonic temperature, and has been interpreted as a sign for electromagnetic interactions between baryons and dark matter.
This hypothesis has
a striking prediction: large temperature anisotropies sourced by the velocity-dependent cooling of the baryons.
However, in order to remain consistent with the rest of cosmological observations, only part of the  dark matter is allowed to be charged, and thus interactive.
Here we compute, for the first time, the 21-cm fluctuations caused by a charged subcomponent of the dark matter, including both the pre- and post-recombination evolution of all fluids.
We find that, 
for the same parameters that can explain the anomalous 21-cm absorption signal,
any percent-level fraction of charged dark matter would source novel 21-cm fluctuations with a unique acoustic spectrum, and with an amplitude above any other known effects.
These fluctuations are uncorrelated with the usual adiabatic anisotropies, and would be observable at high significance with interferometers such as LOFAR and HERA, thus providing a novel probe of dark matter at cosmic dawn.
\end{abstract}

\maketitle

The cosmic dawn---the period of the first stellar formation~\cite{loeb2013first}---provides one of the most extreme environments in our Universe, as it is the era with the slowest dark matter-baryon relative motion.
This epoch is, thus, ideal to search for interactions mediated by nearly massless fields~\cite{Tashiro:2014tsa,Munoz:2015bca,Barkana:2018lgd}.
This possibility has received recent interest, as the EDGES collaboration reported a global-signal measurement of 21-cm absorption at $z=17$ which is a factor of two deeper than expected~\cite{Bowman:2018yin}, and would arise naturally if the baryons had a lower temperature than in the standard cosmological model.
In Ref.~\cite{Munoz:2018pzp} it was shown that
the necessary baryonic cooling can be explained if a fraction of the dark matter (DM) is electrically charged, which partially thermalizes with baryons during the cosmic dawn.
In this \emph{Letter} we will calculate the unique signature of a charged subcomponent of the dark matter in the  21-cm fluctuations.

In addition to baryonic cooling, the cosmic dawn is an excellent probe of energy injection~\cite{Clark:2018ghm, Liu:2018uzy,DAmico:2018sxd,Mitridate:2018iag,Hektor:2018qqw,Cheung:2018vww} and of low-frequency spectral distortions of the cosmic microwave background (CMB), the latter of which has been proposed as a possible explanation of the EDGES signal~\cite{Feng:2018rje,Fraser:2018acy,Pospelov:2018kdh,Ewall-Wice:2018bzf}.
In contrast to this mechanism, interactions between charged DM and baryons predict a new kind of temperature fluctuations, sourced by the velocity modulation of the baryonic cooling, as opposed to the usual adiabatic anisotropies~\cite{Munoz:2015bca}.
Therefore, in addition to potentially confirming the anomalous EDGES detection, measurements of the 21-cm power spectrum will be able to shed light on the properties of dark matter.

In the global-signal analysis of Ref.~\cite{Munoz:2018pzp}, reducing the abundance of charged particles ($\chi$), by lowering the fraction $f_{\rm dm}$ of the total DM they constitute, could be compensated by increasing their charge.
However, this reduces the effect of velocities on the baryonic cooling in two ways. 
First, for small $f_{\rm dm}$ the $\chi$ fluid can, through interactions with baryons, acquire a thermal sound speed that dominates over the $\chi$-baryon ($\chi$-b) bulk relative velocity~\cite{Munoz:2018pzp}.
Second, the  $\chi$-b velocity can be damped before it is frozen at the end of recombination.
Thus, a joint measurement of the global 21-cm temperature, and its fluctuations, can break the apparent degeneracy between the abundance of $\chi$ particles and their charge~\cite{Munoz:2018pzp}, and be complementary to CMB searches for interactive DM~\cite{Dvorkin:2013cea,Xu:2018efh,Boddy:2018kfv,Gluscevic:2017ywp}.

In the standard cosmological model with pure cold dark matter (cdm), 
only baryons are affected by radiation pressure until recombination.
This sources a relative velocity between them and the DM~\cite{Tseliakhovich:2010bj}, which has a root mean square (rms) value of $v_{\rm rms}^{\rm cdm}\approx 29$ km s$^{-1}$ at kinematic decoupling (redshift $z_{\rm kin}\approx 1010$), after which it redshifts like $(1+z)$, as baryons start falling into the dark-matter gravitational potential.
The shape of its power spectrum shows strong acoustic oscillations, as it is sourced by baryon-photon interactions.

However, the $\chi$ fluid is not inert, so fluctuations on the $\chi$-b velocity might be damped prior to $z_{\rm kin}$, as both fluids couple kinematically.
To account for this effect, we solve the fluid equations of the $\chi$, baryon, and cdm fluids, where the charged dark matter composes a fraction $f_{\rm dm}$ of the total dark matter (d) and the rest is neutral cdm. 
We use a modified version of the publicly available code {\tt CAMB}~\cite{Lewis:1999bs} that includes scattering of baryons with a subcomponent of the DM~\cite{Dvorkin:2013cea,Xu:2018efh}, which allows us to consistently calculate the fluctuations in the $\chi$-b relative velocity accounting for their interactions.
Throughout this work we will adopt fiducial cosmological parameters consistent with Ref.~\cite{Ade:2015xua}, of (physical) baryon and total-DM densities $\omega_b=0.022$ and $\omega_d=0.12$, and $h=0.67$.

\begin{figure}[hbtp!]
	\includegraphics[width=0.5\textwidth]{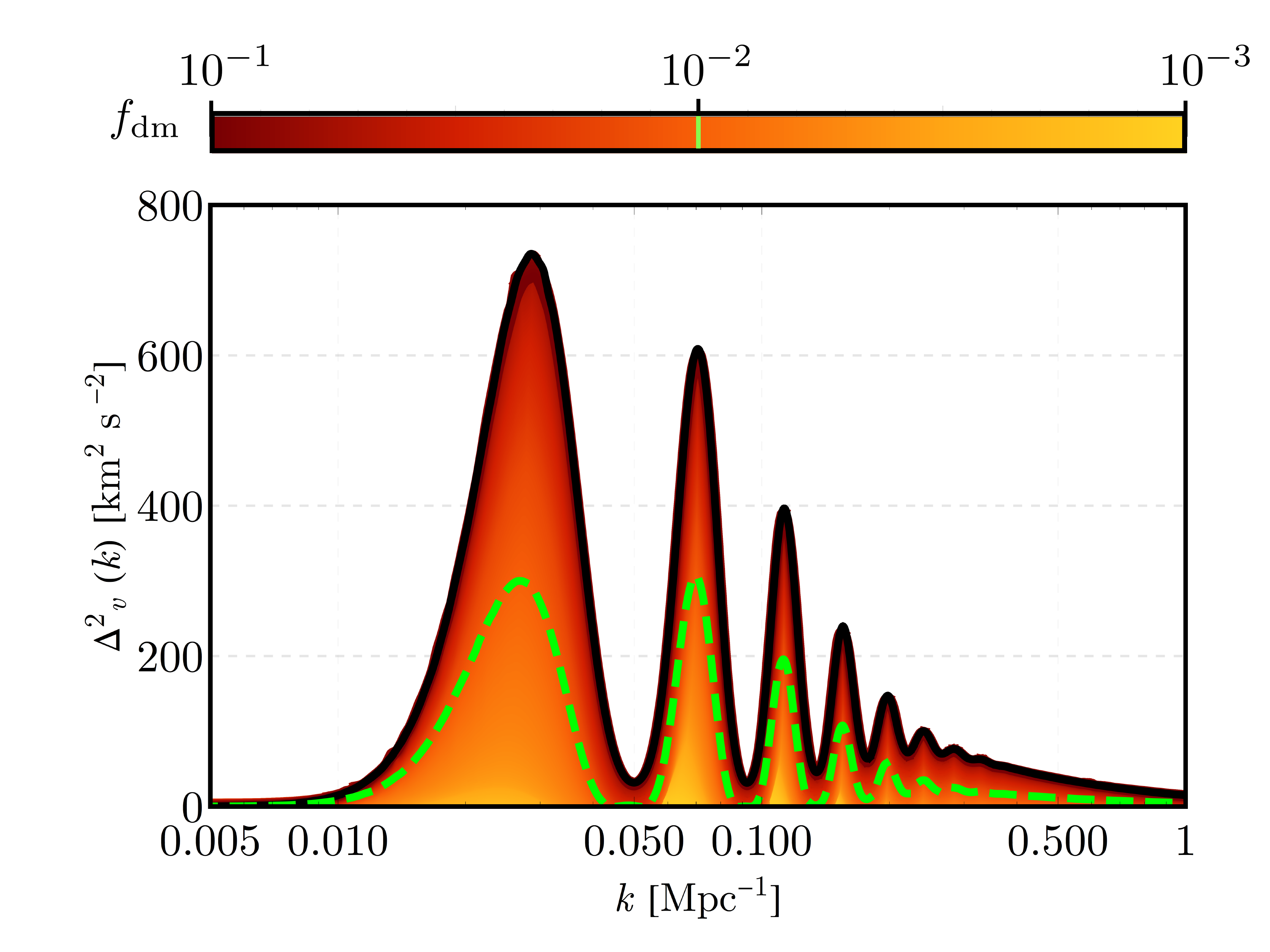}
	\caption{Amplitude of fluctuations of the $\chi$-baryon relative velocity, as a function of comoving wavenumber $k$.
	In black we show the result without interactions, and in color scale from red to yellow we show smaller fractions $f_{\rm dm}$ of charged dark matter, with increasing charges given by the Eq.~\eqref{eq:epsilon}.
	To guide the eye, the green-dashed line shows the $f_{\rm dm}=0.01$ case.	
	}
	\label{fig:Dvsq_k}
\end{figure}

We compute the $\chi$-b relative velocity as
\be
v_{\chi b} =- \dfrac{1}{i k} H(z) \dfrac{d}{dz} (\delta_b - \delta_\chi),
\ee
at $z_{\rm kin}$, where $H$ is the Hubble parameter, and $\delta_{b(\chi)}$ is the baryon ($\chi$) overdensity, and define its power spectrum $P_v$ through
\be
\VEV{v_{\chi b} (\mathbf k) v_{\chi b} (\mathbf k')}=(2\pi)^3\delta_D(\mathbf k+\mathbf k') P_v(k),
\ee
and its amplitude of fluctuations as $\Delta_v^2(k) = k^3 P_v(k)/(2\pi^2)$. We show this quantity in Fig.~\ref{fig:Dvsq_k} for different values of $f_{\rm dm}$ and charge, where the latter is increased for decreasing $f_{\rm dm}$, so as to explain the anomalous 21-cm depth as in Ref.~\cite{Munoz:2018pzp} (see also~\cite{Barkana:2018qrx,Slatyer:2018aqg,Mahdawi:2018euy}), by the approximate relation
\be
\epsilon = 6 \times 10^{-7} \, \left(\dfrac{m_\chi}{\rm MeV} \right) \left(\dfrac{f_{\rm dm}}{10^{-2}}\right)^{-3/4},
\label{eq:epsilon}
\ee
valid for fractions $f_{\rm dm}\lesssim0.1$.
Here $m_\chi$ is the charged-DM mass, which we can set to any value $m_\chi\ll 6\, f_{\rm dm} \rm GeV$ without loss of generality, and the minicharge $\epsilon$ is in units of the electron charge.
We ignore the effect of heating by absorption of CMB photons~\cite{Venumadhav:2018uwn}, as it only changes the necessary charges at the $\sim 10\%$ level and does not induce any important 21-cm fluctuations.
This Figure shows how reducing the fraction $f_{\rm dm}$ of charged dark matter (increasing its charge), results in a damped relative-velocity power spectrum with baryons, as the two fluids couple.
We find that for $f_{\rm dm}\gtrsim0.03$ there is no significant damping relative to the noninteracting case, whereas for  $f_{\rm dm}\lesssim10^{-3}$ there is essentially no $\chi$-b relative velocity, as these two fluids are kinematically coupled. 
Values in between see a significant reduction of the rms velocity, which we calculate as
\be
v_{\rm rms} \equiv \left[ \int\!\!\! \dfrac{d^3k}{(2\pi)^3} P_v(k)\right]^{1/2} \!\!\!\!\!\! \approx \left[0.6 + 0.3 \, \log\left(\dfrac{f_{\rm dm}}{10^{-2}}\right) \right] v_{\rm rms}^{\rm cdm},
\ee
where the last approximation is a fit
valid for $10^{-3} \lesssim f_{\rm dm} \lesssim 0.03$.
We will assume that the resulting probability distribution function (PDF) $\mathcal P$ for each component of the $\chi$-b relative velocity remains Gaussian, and thus in all cases $\mathcal P(v_{\chi b})$ is given by a Maxwell-Boltzmann distribution, with rms velocity $v_{\rm rms}$.
For the rest of this work we will focus on values of the charge given by Eq.~\eqref{eq:epsilon}, although we note that higher charges can still explain the enhanced 21-cm absorption, while producing additional damping of the $\chi$-b velocity prior to recombination.

We will ignore interactions between $\chi$ particles and the rest of the DM, since thermalization of those two fluids would require self-interaction rates orders of magnitude above contemporary limits for heavy mediators~\cite{Harvey:2015hha}, or at the edge of current constraints for light mediators~\cite{Agrawal:2016quu}.
Additionally, we note that a small fraction of the DM is allowed to interact (and even equilibrate) with baryons at recombination, as long as the apparent increase in $\omega_b$ is within observational limits.
This constraint corresponds to $f_{\rm dm}<0.6\%$~\cite{Dolgov:2013una,dePutter:2018xte} for fully coupled fluids, although for the charges in Eq.~\eqref{eq:epsilon} it would be less strict.

We now simultaneously solve for the post-recombination evolution of the $\chi$ and baryonic temperatures, their relative velocity, and the free-electron fraction.
The inclusion of $\chi$-b relative velocities has two main effects.
First, it provides a reservoir of kinetic energy, which interactions can convert into thermal energy.
However, for the $\chi$ mass range we are interested in, this effect is subdominant.
Second, the overall velocity modulates the interaction rate, as the cross section is strongly dependent on velocities.
We follow the method described in Ref.~\cite{Munoz:2015bca} (see also~\cite{AliHaimoud:2010dx,Ma:1995ey}), with the modifications due to charged DM included in Ref.~\cite{Munoz:2018pzp}.

The (baryonic) gas temperature $T_g$ at high redshifts is only observable through its effect on the 21-cm brightness temperature $T_{21}$, which can be written as~\cite{McQuinn:2005hk,Pritchard:2008da}
\be
T_{21} = 36 \, {\rm mK} \, x_{\rm HI} 
 \left(\dfrac{1+z}{18}\right)^{1/2}  \left( 1- \dfrac{T_{\rm cmb}}{T_S}\right),
\label{eq:T21}
\ee
where $x_{\rm HI}$ is the atomic-hydrogen fraction (which we will set to unity), and $T_S$ and $T_{\rm cmb}$ are the spin and CMB temperatures, respectively.
For the sake of simplicity,
we will assume saturated Lyman-$\alpha$ coupling ($T_S=T_g$), and no X-ray heating.
These two factors would change dramatically both at low redshifts, where X-rays would heat up the gas, and at high reshifts, where Lyman-$\alpha$ coupling is not fully efficient. 
Thus, we will focus on the redshift range $17<z<20$, corresponding to the EDGES data~\cite{Bowman:2018yin}.
We note, however, that our 21-cm fluctuations $\delta_{T_{21}}^{(\rm this\, work)}\equiv \delta T_{21}/T_{21}$ can be rescaled to any subsaturated Lyman-$\alpha$ case simply as~\cite{Pritchard:2008da}
\be
\delta_{T_{21}}^{(\rm new)} = \left( \dfrac{T_g^{(\rm this\, work)}-T_{\rm cmb}}{T_S^{(\rm new)}-T_{\rm cmb}} \right)\delta_{T_{21}}^{(\rm this\, work)},
\ee
where $T_S^{(\rm new)}$ is the new spin temperature, and $T_g^{(\rm this\, work)}=4$ K at $z=17$, scaling roughly as $(1+z)^2$, and we have ignored Lyman-$\alpha$ fluctuations, as these are uncorrelated with the relative velocity.
This would not be the case once X-ray heating starts to dominate, although of course  any 21-cm fluctuations induced by cooling of the baryons would be quickly washed away as soon as heating is important.

We are only computing the 21-cm fluctuations induced by the velocity-dependent cooling of the baryons, as those are a unique signature of charged DM. 
Nonetheless, these fluctuations are uncorrelated with density perturbations to first order~\cite{Dalal:2010yt}, as they trace $v_{\chi,b}^2$, so one can linearly add the 21-cm power spectrum we calculate to that originating from usual anisotropies (which can be found, for instance, with the publicly available code {\tt 21cmFAST}~\cite{Mesinger:2010ne}), to find the total 21-cm power spectrum.
We will see that the shapes of these two power spectra are rather different, so we will focus on the velocity-induced fluctuations for the rest of this work.

\begin{figure}[hbtp!]
	\hspace{-0mm}
	\includegraphics[width=0.5\textwidth]{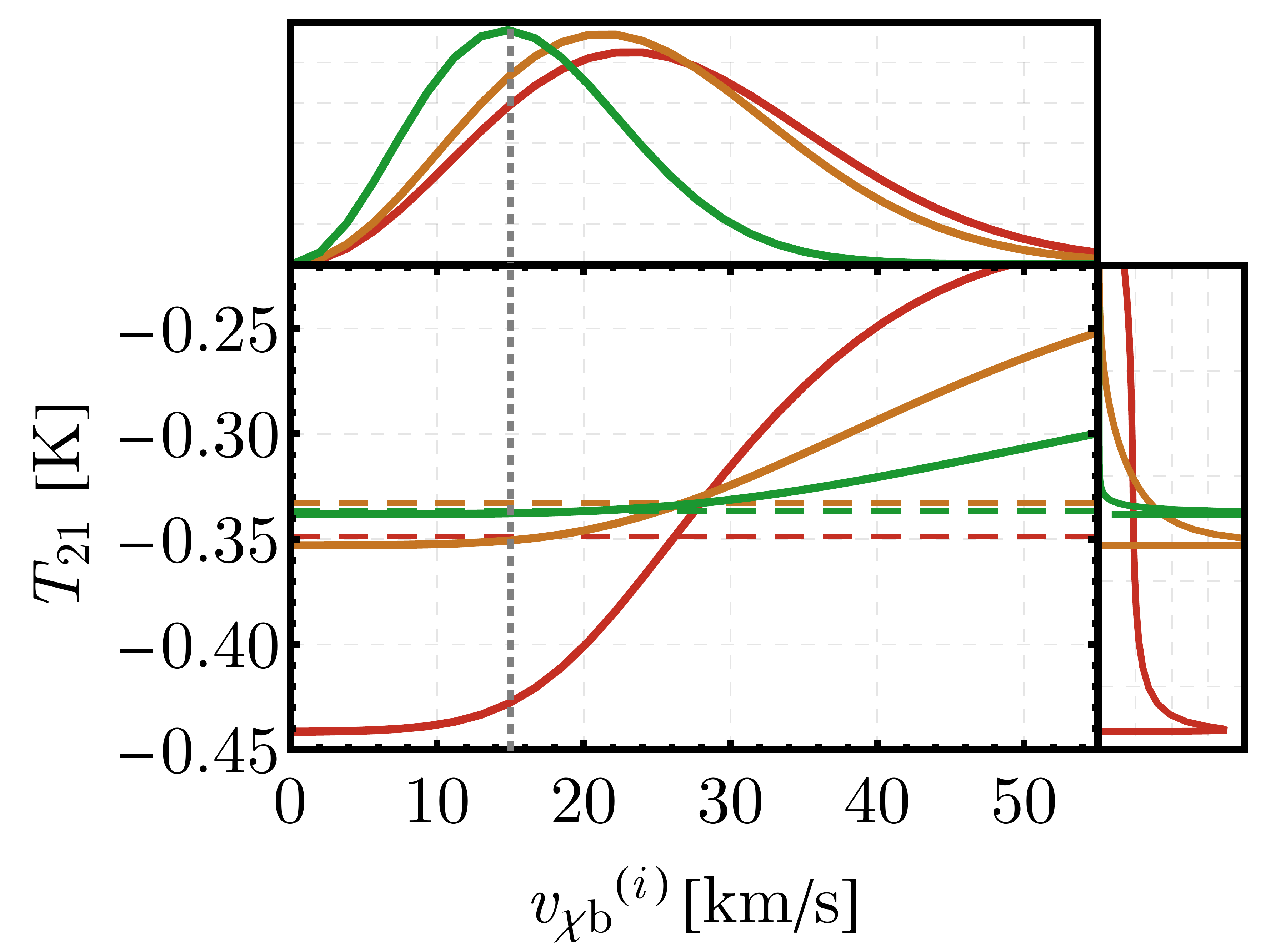}
	\caption{The 21-cm brightness temperature at $z=17$, assuming negligible X-ray heating and full Lyman-$\alpha$ coupling, as a function of the (initial) $\chi$-baryon relative velocity at $z=z_{\rm kin}$.
	The  red, orange, and green lines correspond to  $f_{\rm dm}= \{0.1,$  $0.03$, $0.01\}$, with charges $\epsilon/m_\chi= \{0.8,$ 2, 5$\}\times 10^{-7} \, \rm MeV^{-1}$ (and $m_\chi\ll 6 f_{\rm dm} \, \rm GeV$), respectively.
	Dashed horizontal lines represent the sky-averaged $T_{21}$, and
	we show the (unnormalized) PDFs of the relative velocity (upper panel) and 21-cm temperature (right panel) for each value of $f_{\rm dm}$.
	The dotted vertical line represents the value of the relative velocity that would be subthermal at $z\leq150$, in the absence of interactions.
	}
	\label{fig:T21_v}
\end{figure}

Fig.~\ref{fig:T21_v} shows the 21-cm temperature at $z=17$, where higher velocities provide less cooling (as the interaction rate is suppressed), and thus cause a shallower absorption in $T_{21}$.
Additionally, low relative velocities yield very similar temperatures, as the motion is dominated by the thermal velocity---which after $z\sim 150$ is roughly half of the (undamped) DM-baryon rms velocity.
We also show the PDF of the $\chi$-b velocity (including the effect of damping prior to recombination), as well as the inferred PDF for $T_{21}$ for each of the cases.
This last PDF is clearly not Gaussian, and presents a peak at a minimum temperature $T_{21}^{\rm min}=T_{21}(v_{\chi b}^{(i)}=0)$, and a tail extending to larger values. 
Moreover, the width of this PDF (and thus the amplitude of the 21-cm fluctuations) grows for larger values of $f_{\rm dm}$, even for similar averaged $T_{21}$, as expected.

We now find the 21-cm fluctuations for different values of the fraction $f_{\rm dm}$ of charged dark matter, again fixing the charge to fit the EDGES data as in Ref.~\cite{Munoz:2018pzp}.
Given the 21-cm temperature $T_{21}(v_{\chi b})$, we can compute its correlation function as~\cite{Dalal:2010yt}
\be
\VEV{T_{21}(0) T_{21}(\mathbf x)} = \int d^3 \mathbf v_0 d^3 \mathbf v_x \mathcal P(\mathbf v_0, \mathbf v_x) T_{21}(v_0) T_{21} (v_x),
\label{eq:T21corr}
\ee
where $\mathbf v_0$ and $\mathbf v_x$ are the relative velocities at the origin and at position $\mathbf x$, and $\mathcal P(\mathbf v_0, \mathbf v_x)$ is their joint PDF, given by a six-dimensional Gaussian~\cite{Dalal:2010yt,Ali-Haimoud:2013hpa}.
This distribution encodes the correlations between the parallel and perpendicular components of the velocity at different points, given by~\cite{Dalal:2010yt,Ali-Haimoud:2013hpa}
\begin{subequations}
	\begin{align}
	\psi_{||}(x) =& \dfrac{1}{v_{\rm rms}^2}\int \dfrac{d^3k}{(2\pi)^3} P_v(k) \left[ j_0(kx) - 2 j_2 (kx) \right], \\
	\psi_{\perp}(x) =& \dfrac{1}{v_{\rm rms}^2} \int \dfrac{d^3k}{(2\pi)^3} P_v(k) \left[ j_0(kx) + \, j_2 (kx) \right],
	\end{align}
\end{subequations}
where $j_{\ell}$ are the spherical Bessel functions of the first kind, and $P_v(k)$ is computed for each $f_{\rm dm}$ as in Fig.~\ref{fig:Dvsq_k}.
The two-point function of 21-cm fluctuations is then found as $\xi_{T_{21}} (x) = \VEV{T_{21}(0) T_{21}(x)}/\VEV{T_{21}}^2 - 1$,
from where we obtain the sought-after  power spectrum of 21-cm fluctuations to be
\be
P_{T_{21}}(k) = (4\pi) \VEV{T_{21}}^2 \int_0^\infty d x\,x^2 \xi_{T_{21}} (x) j_0(k x).
\ee
For convenience we define the amplitude of 21-cm perturbations as $\Delta_{T_{21}}^2(k) = k^3 P_{T_{21}}(k)/(2\pi^2)$.

\begin{figure}[t!]
	\hspace{-0mm}
	\includegraphics[width=0.48\textwidth]{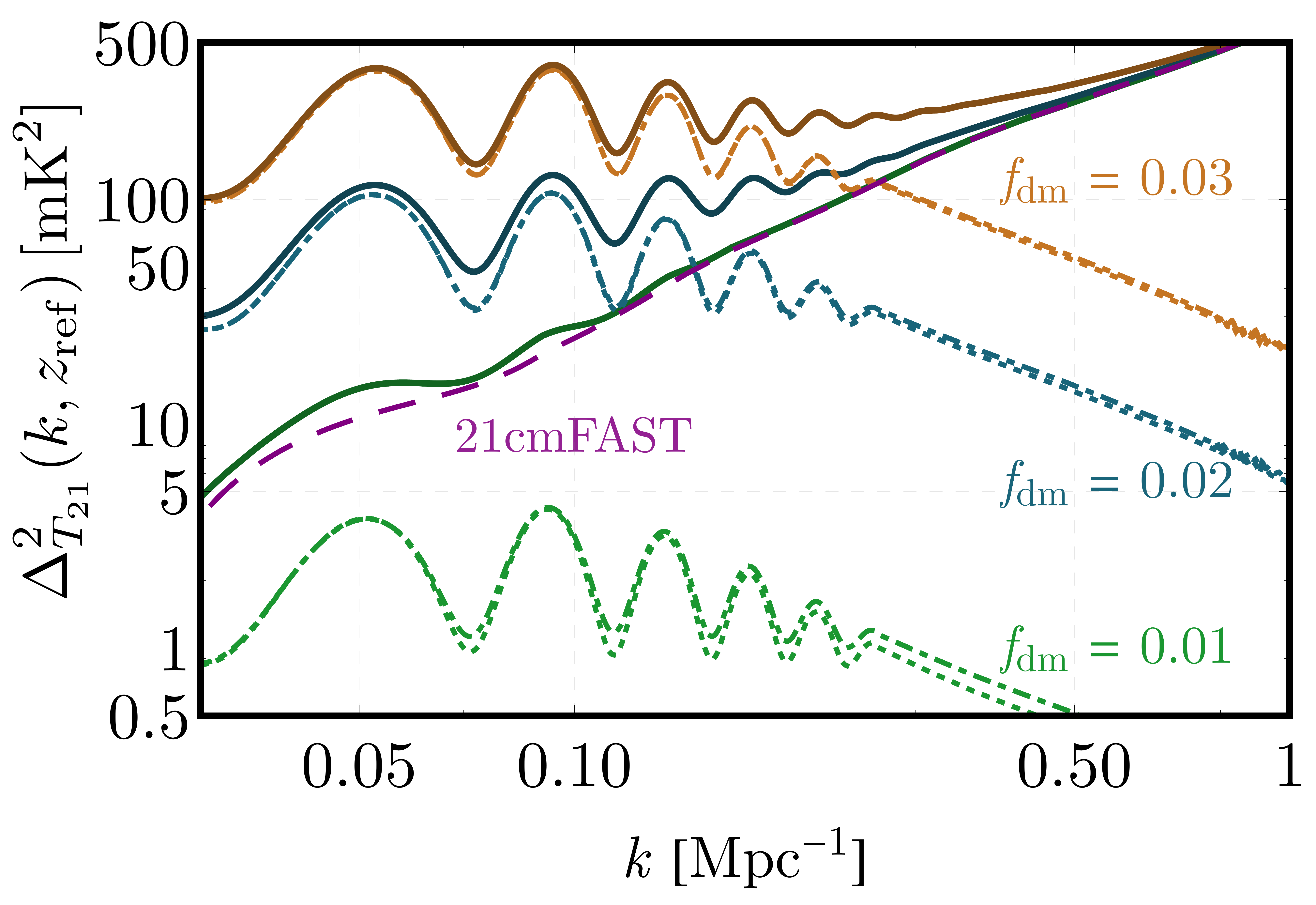}
	\caption{Amplitude of the 21-cm temperature power spectrum as a function of $k$, at $z_{\rm ref}=17$. 
	Solid colored line represent the total power spectrum for each value of $f_{\rm dm}$, obtained by linearly adding the standard contribution from {\tt 21cmFAST} (in dashed purple), and the contribution from charged DM, with charges given by Eq.~\eqref{eq:epsilon}, in (dash-dotted lines).
	The dotted lines show the approximation of Eq.~\eqref{eq:P_T_bias}, for each charged-DM case.
	}
	\label{fig:Dsq_k}
\end{figure}

We show the resulting $\Delta_{T_{21}}^2(k)$ at $z=17$ in Fig.~\ref{fig:Dsq_k}, along with the standard ``out-of-the-box" output from {\tt 21cmFAST}, shown for illustrative purposes only.
We find, in all cases, a strong acoustic signature in the DM-induced 21-cm fluctuations, peaking at $k\sim0.1$ Mpc$^{-1}$ and decaying for larger and smaller wavenumbers.
This is to be compared with the standard prediction from {\tt 21cmFAST}, which does not show important acoustic oscillations.
It is in this sense that charged DM produces unique anisotropies,  always tracing  $\Delta^2_{v^2}\equiv k^3 P_{v^2}(k)/(2\pi^2)$ (i.e., the amplitude of fluctuations of $v_{\chi,b}^2$), which cannot be easily mimicked by any parameter combination within {\tt 21cmFAST}.

In order to develop some analytical understanding of our result, we follow Ref.~\cite{Dalal:2010yt} in writing the approximate relation
\be
\Delta^2_{T_{21}} (k,z) \approx  b_{T_{21}}^2(z) \Delta^2_{v^2}(k),
\label{eq:P_T_bias}
\ee
where the effective bias $b_{T_{21}}$ is defined through
\be
b_{T_{21}}^2 
= \dfrac{\VEV{T_{21}^2}-\VEV{T_{21}}^2}{\VEV{v_{\chi b}^4}-v_{\rm rms}^4},
\label{eq:b21}
\ee
where $\Delta^2_{v^2}$ and  $v_{\rm rms}$ are evaluated for each $f_{\rm dm}$ at $z=z_{\rm kin}$. 
We show this approximation in Fig.~\ref{fig:Dsq_k}, where  it is evident that it is, in fact, close to the full calculation at all scales in the problem and, thus, that it provides a universal fit to the fluctuations induced by charged DM.
We find the 21-cm rms fluctuations at any redshift simply as $T_{21}^{\rm rms}(z) = b_{T_{21}}(z) \, v_{\rm rms}^2$, which for fractions $f_{\rm dm}=\{0.1, 0.03, 0.02, 0.01\}$,  yields $T_{21}^{\rm rms}=\{110, 30, 16, 3\}$ mK at $z=17$.
Additionally, we find that the effective bias function can be approximated as $b_{T_{21}}(z) \approx 4.4 \, {\rm mK\,  km^{-2}\,  s^2} \, f_{\rm dm}^{4/3} \, [18/(1+z)]^2$.

We have, thus far, focused on the 21-cm fluctuations induced by the velocity-dependent cooling of the baryons, as we argued they follow relative-velocity fluctuations, and thus have a unique shape.
Nonetheless, there are known astrophysical effects that can imprint $\Delta^2_{v^2}$ in the 21-cm fluctuations during cosmic dawn.
Chiefly, there is a velocity-dependent suppression in the abundance~\cite{Tseliakhovich:2010bj,Naoz:2011if} and gas content~\cite{Dalal:2010yt,Naoz:2012fr} of small haloes, as well as in their star-formation efficiency~\cite{Maio:2010qi,Stacy,Fialkov:2011iw}.
These effects cause 21-cm fluctuations that depend on the value of the total DM-baryon velocity $v_{d b}$, which does not suffer significant damping. 
At high redshifts, anisotropic Lyman-$\alpha$ pumping was estimated to cause a 21-cm rms temperature of $\sim 2$ mK~\cite{Dalal:2010yt}, whereas at lower redshifts 21-cm fluctuations can also be sourced by inhomogeneities in the X-ray emission---and thus gas heating---with a total estimated amplitude of 10 mK~\cite{Visbal:2012aw}.
We note, however, that our calculations show that for $f_{\rm dm} \gtrsim 0.02$ the 21-cm fluctuations induced by charged dark matter are larger than any of these effects, and
and would dominate the signal at large scales (although of course not at smaller scales, where relative velocities are coherent).
Additionally, any velocity-induced fluctuations in $x_\alpha$ would only increase the size of the signal we calculate, as a high relative velocity produces both fewer stars (and thus fewer Lyman-$\alpha$ photons) and less cooling, resulting in an overall shallower 21-cm absorption, which shows that the two effects act in the same direction.
For comparison, we estimate that for $f_{\rm dm}=1$, the charges that can explain the EDGES signal also produce 21-cm fluctuations with $T_{21}^{\rm rms} \approx 500$ mK, in agreement with the simulations of Ref.~\cite{Fialkov:2018xre}, which would be easily observable with radio interferometers~\cite{1804.00515}.
Nonetheless, the $f_{\rm dm}=1$ case is excluded jointly by stellar-cooling constraints, fifth-force experiments, and Galactic considerations~\cite{Munoz:2018pzp,Knapen:2017xzo}.

We will now estimate the observability of the signal with two upcoming 21-cm interferometers, LOFAR\footnote{\url{ http://www.lofar.org/}} and HERA\footnote{\url{http://reionization.org/}}, and
calculate their sensitivity following Ref.~\cite{Tegmark:2008au}.
For LOFAR, we take the NL Inner configuration of the LBA~\cite{vanHaarlem:2013dsa}.
For HERA we consider the 320-antenna core, with a total collecting area of $A_{\rm coll}=49260$ m$^2$~\cite{DeBoer:2016tnn}, which we take to be frequency independent.
For both arrays we assume a sky coverage of 1440 deg$^2$, as that is the HERA field, corresponding to $f_{\rm sky}\approx 3.5\%$, as well as a system temperature $T_{\rm sys} (\nu)=100 + 120 (\nu/150 \, \rm MHz)^{-2.55}$ K~\cite{DeBoer:2016tnn,Pober2013,Pober:2013jna}, and an observation time of 1080 hrs (45 days).
We will estimate sensitivities for $f_{\rm dm}$ as low as 1\%, as
it might be possible to distinguish DM-induced cooling from astrophysical effects, since perturbations in the Lyman-$\alpha$ pumping would show suppression at larger scales than cooling~\cite{Dalal:2010yt}, and would be most important at early times; whereas anisotropies in X-ray heating dominate at later times~\cite{Visbal:2012aw,McQuinn:2012rt}, and their velocity dependence is subdominant if strong feedback is present~\cite{Fialkov:2012su}. 

\begin{figure}[hbtp!]
	\hspace{-5mm}
	\includegraphics[width=0.5\textwidth]{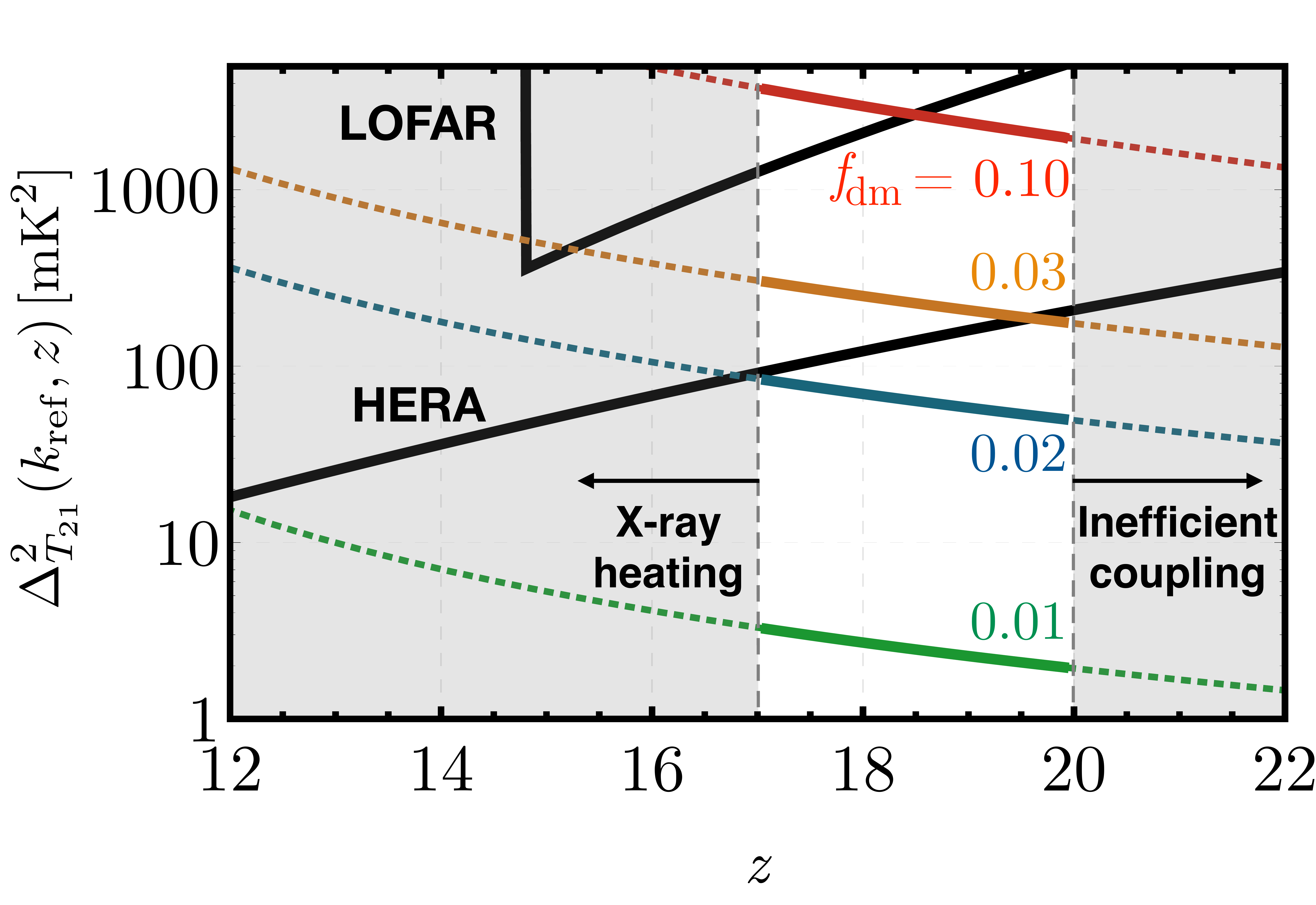}
	\caption{Amplitude of the 21-cm temperature power spectrum induced by different fractions of charged DM, with charges given by Eq.~\eqref{eq:epsilon}, as a function of redshift, at a reference scale $k_{\rm ref}=0.2 \, h$ Mpc$^{-1}$.
		We shade out the regions $z>20$, due to the inefficient Lyman-$\alpha$ coupling, and $z<17$, where X-ray heating is expected to be important.
		We also show the sensitivity of HERA and LOFAR with 1080 hrs of observation, as black lines.
	}
	\label{fig:Dsq_z}
\end{figure}

In Fig.~\ref{fig:Dsq_z}  we show, as a function of redshift, the amplitude of the 21-cm fluctuations sourced by charged DM at $k=0.2\, h\,$Mpc$^{-1}$ for different values of $f_{\rm dm}$, as well as the LOFAR and HERA instrumental noises.
We do not include any other source of fluctuations to remain conservative, as those are not unique to charged DM. If desired, they can be linearly added to the amplitude shown in Fig.~\ref{fig:Dsq_z}.
From a simplistic signal-to-noise ratio (SNR) analysis we find that, integrating over scales $0.15\, h\, {\rm Mpc^{-1}} < k < 0.5\, h\, {\rm Mpc^{-1}}$ to simulate foreground removal~\cite{Ali:2015uua}, and considering the redshift range $z=17$ to $18$, LOFAR and HERA should be sensitive to $f_{\rm dm}\geq 0.02$ with SNRs larger than 15 and 150, respectively, and HERA would probe down to $f_{\rm dm}=0.01$ with a SNR of 10, although we emphasize that separation from other velocity-dependent astrophysical sources of 21-cm fluctuations might be required for any detection below $f_{\rm dm}<0.02$.

In summary, we have computed, for the first time, the 21-cm fluctuations induced by partially charged dark matter, with the necessary charges to explain the EDGES detection~\cite{Bowman:2018yin,Munoz:2018pzp}.
We have shown that, in all cases, the induced 21-cm fluctuations approximately trace $\Delta^2_{v^2}$, with an amplitude growing with the fraction $f_{\rm dm}$ of charged DM.
Since other sources of anisotropies are uncorrelated with these fluctuations, one can probe charged DM by simply searching for a nonzero $\Delta^2_{v^2}$ in 21-cm data.
We find that for fractions of charged dark matter above a percent  these fluctuations are within the reach of interferometers such as HERA and LOFAR.
Therefore, these experiments will be sensitive to even small traces of charged dark matter at cosmic dawn and, thus, will confirm or severely constrain whether we have detected interactions of dark matter with baryons.

\acknowledgements

We wish to thank Azadeh Moradinezhad for discussions, Marc Kamionkowski for comments on a previous version of this manuscript, and the anonymous referees for their feedback.
This work was supported by the Dean's Competitive Fund for Promising Scholarship at Harvard University, 
and by the Black Hole Iniative, which is funded by a grant from the John Templeton Foundation.
\newline

\bibliography{Dmb_bib}{}
\bibliographystyle{bibpreferences}

\end{document}